\documentclass[letterpaper]{article} % DO NOT CHANGE THIS
\usepackage{aaai2026}  % DO NOT CHANGE THIS
\usepackage{times}  % DO NOT CHANGE THIS
\usepackage{helvet}  % DO NOT CHANGE THIS
\usepackage{courier}  % DO NOT CHANGE THIS
\usepackage[hyphens]{url}  % DO NOT CHANGE THIS
\usepackage{graphicx} % DO NOT CHANGE THIS
\usepackage{natbib}  % DO NOT CHANGE THIS AND DO NOT ADD ANY OPTIONS TO IT
\usepackage{caption} % DO NOT CHANGE THIS AND DO NOT ADD ANY OPTIONS TO IT
\usepackage{algorithm}
\usepackage{algorithmic}
\usepackage{amsmath}
\usepackage{amssymb}
\usepackage{booktabs}   
\usepackage{pifont}
\usepackage{newfloat}
\usepackage{listings}
\DeclareCaptionStyle{ruled}{labelfont=normalfont,labelsep=colon,strut=off} % DO NOT CHANGE THIS
\floatstyle{ruled}
\newfloat{listing}{tb}{lst}{}
\floatname{listing}{Listing}
\usepackage{helvet}
\usepackage{courier}
\usepackage{xcolor}
\title{CAT-Net: A Cross-Attention Tone Network for Cross-Subject EEG-EMG Fusion Tone Decoding}
\author{
    Yifan Zhuang\textsuperscript{\rm 3},
    Calvin Huang\textsuperscript{\rm 4},
    Zepeng Yu\textsuperscript{\rm 5},
    Yongjie Zou\textsuperscript{\rm 2}\equalcontrib,
    Jiawei Ju\textsuperscript{\rm 1,\rm 2}\equalcontrib
}
\affiliations{
    \textsuperscript{\rm 1}Shanghai Center for Brain Science and Brain-Inspired Technology\\
    \textsuperscript{\rm 2} Lingang Laboratory\\
    \textsuperscript{\rm 3}Sony Interactive Entertainment, United States\\
    \textsuperscript{\rm 4}Amazon.com\\
    \textsuperscript{\rm 5}Independent Researcher\\
    evelyn.zhuang@sony.com, chuang28@stevens.edu, jeffreyzepengyu@ucla.edu, zouyj@lglab.ac.cn, jiawei\_ju@163.com
}

\begin{document}

\maketitle

\begin{abstract}
Brain-computer interface (BCI) speech decoding has emerged as a promising tool for assisting individuals with speech impairments. In this context, the integration of electroencephalography (EEG) and electromyography (EMG) signals offers strong potential for enhancing decoding performance. Mandarin tone classification presents particular challenges, as tonal variations convey distinct meanings even when phonemes remain identical. In this study, we propose a novel cross-subject multimodal BCI decoding framework that fuses EEG and EMG signals to classify four Mandarin tones under both audible and silent speech conditions. Inspired by the cooperative mechanisms of neural and muscular systems in speech production, our neural decoding architecture combines spatial-temporal feature extraction branches with a cross-attention fusion mechanism, enabling informative interaction between modalities. We further incorporate domain-adversarial training to improve cross-subject generalization. We collected 4,800 EEG trials and 4,800 EMG trials from 10 participants using only twenty EEG and five EMG channels, demonstrating the feasibility of minimal-channel decoding. Despite employing lightweight modules, our model outperforms state-of-the-art baselines across all conditions, achieving average classification accuracies of 87.83\% for audible speech and 88.08\% for silent speech. In cross-subject evaluations, it still maintains strong performance with accuracies of 83.27\% and 85.10\% for audible and silent speech, respectively. We further conduct ablation studies to validate the effectiveness of each component. Our findings suggest that tone-level decoding with minimal EEG-EMG channels is feasible and potentially generalizable across subjects, contributing to the development of practical BCI applications.
\end{abstract}

\begin{links}
    \link{Code}{https://github.com/YifanZhuang/CAT-Net}

\end{links}

\section{Introduction}
Brain-computer interface (BCI) speech decoding represents a transformative technology for individuals with speech impairments caused by stroke, ALS, and brainstem injuries \cite{brumberg2010brain}. Among tonal languages, Mandarin Chinese presents unique challenges due to its four distinct tones, which fundamentally alter word meanings despite identical phonemes. The development of effective tone classification systems is crucial for creating practical BCI communication interfaces that can operate effectively in Chinese-speaking populations.

Electroencephalography (EEG) has emerged as a promising non-invasive technique for neural speech decoding, offering excellent temporal resolution and portability advantages \cite{sharon2020neural, dash2020role}. EEG captures neural activities associated with speech planning, phonological processing, and auditory feedback mechanisms. However, extracting Mandarin tones from EEG signals presents significant challenges due to the subtle and spatially distributed nature of pitch-related neural signals. Traditional EEG systems suffer from limited spatial resolution, resulting in high misclassification rates, particularly for perceptually challenging tone pairs such as Tone 2 (rising) and Tone 4 (falling), which share similar pitch contour characteristics and are frequently confused in both neural decoding and perceptual studies \cite{li2021mandarin}.

To address these limitations, researchers have increasingly explored multimodal approaches by integrating surface electromyography (sEMG) signals from facial and jaw muscles during both audible and silent speech conditions \cite{rojas2022recognition}. EMG provides complementary articulatory information about muscle activation patterns that directly correlate with tone production mechanisms. The activation of specific muscle groups, including the zygomatic and masseter muscles, produces distinct patterns corresponding to different pitch contours in Mandarin tones. By combining EEG and EMG signals, researchers can simultaneously monitor neural speech intentions and peripheral articulatory execution, potentially improving decoding accuracy across both overt and covert speech conditions. In recent studies, Li et al. demonstrated that fusing EEG and EMG signals can lead to more robust silent speech recognition systems \cite{li2023hybrid}.

Despite promising results, current EEG-EMG fusion systems face several critical limitations. First, existing approaches typically require high-density electrode configurations with numerous EEG channels and multiple EMG sensors, making them impractical for daily use due to setup complexity and user discomfort \cite{zhang2023sparsity}. Second, most current systems lack effective mechanisms for capturing the intricate interactions between neural and muscular signals, often relying on simple concatenation or basic fusion strategies. Third, and most importantly, these systems demonstrate poor generalization capabilities across different users, with performance degrading substantially when applied to unseen subjects due to anatomical differences, electrode impedance variations, and individual cortical activation patterns.

The cross-subject generalization challenge is particularly critical for practical BCI applications, as clinical systems must function reliably across diverse user populations without requiring extensive individual calibration \cite{chen2025decoding, lee2021Decoding}. Current approaches show significant performance drops when transitioning from within-subject to cross-subject evaluation scenarios, highlighting the need for more robust domain adaptation strategies.

To address these challenges, we introduce CAT-Net, [C]ross-[A]ttention [T]one Net, a novel multimodal fusion framework that integrates EEG and EMG signals for the classification of Mandarin tones. Our approach makes several key contributions: (1) We develop a cross-attention mechanism inspired by Transformer architectures \cite{vaswani2017attention} that enables sophisticated bidirectional interaction between EEG and EMG modalities, moving beyond simple concatenation to capture complex neural-muscular coordination patterns; (2) We demonstrate effective tone classification using a minimal channel configuration of only 20 EEG channels and 5 EMG channels, selected through attention-based channel importance analysis; (3) We incorporate domain adaptation techniques  \cite{ganin2016domain} to enhance cross-subject generalization capabilities; (4) We provide comprehensive evaluation across both audible and silent speech conditions, demonstrating the system's versatility for different communication scenarios, assisting individuals with hearing or speech impairments.

\section{Related Work}

\subsection{Neural Speech Decoding}
EEGNet and DeepConvNet have been widely used as baseline architectures for EEG-based speech decoding tasks due to their lightweight design and strong feature extraction capabilities \cite{lawhern2018eegnet, Schirrmeister2017DeepConvNet}. These models have demonstrated promising results in brain-computer interface (BCI) applications, particularly for assisting individuals with speech impairments \cite{brumberg2010brain}. For Mandarin tone classification, Li et al. \cite{li2021mandarin} achieved 42.9\% accuracy using Riemannian manifold features, while Wang et al. \cite{wang2023cross} improved the accuracy to 68\% with end-to-end CNNs. However, EEG-only approaches suffer from limited spatial resolution and high artifact susceptibility, particularly when distinguishing similar tones like Tone 2 and Tone 4.

EMG signals provide complementary articulatory information that directly correlates with speech production \cite{bharali2024enabling}. Wu et al. achieved 90.76\% accuracy on Mandarin phrase classification using parallel Inception CNNs with EMG \cite{wu2022novel}. Additionally, Janke and Diener demonstrated that facial surface electromyographic (sEMG) signals can be directly transformed into audible speech, enabling silent speech communication while preserving critical paralinguistic cues  \cite{Janke2017EMG}. These findings highlight the potential of EEG and EMG as viable modalities for decoding speech, especially in scenarios where traditional acoustic signals are unavailable or impaired.

\subsection{Multimodal Fusion and Attention Mechanisms}
Traditional multimodal fusion employs concatenation or late fusion, which fails to capture temporal dependencies and cross-modal interactions during speech production. Saha and Fels \cite{saha2019hierarchical} demonstrated that attention-based feature selection significantly improves EEG-based imagined speech decoding, achieving 23.45\% improvement over baselines through hierarchical attention mechanisms on joint variability matrices. Li et al. \cite{li2023hybrid} trained sequence-to-sequence decoders that transform the combined EEG- sEMG signals collected during silent speech into audible speech, and resulting audio transcripts showed a mean character-error rate (CER) of 7.22\% across eight speakers. However, these approaches either focus on single-modality self-attention mechanisms or employ simple concatenation-based fusion, without capturing the bidirectional interactions between multiple inputs.

Cross-attention mechanisms, inspired by Transformers \cite{vaswani2017attention}, have shown success in capturing bidirectional interactions between modalities. Tang et al. \cite{tang2024wearable} applied cross-attention to EMG and accelerometer data for tremor classification, outperforming single-modality baselines. Speech-related applications pose unique challenges that demand advanced attention mechanisms to capture neural-muscular interactions. Our cross-attention framework enables EEG and EMG to dynamically attend to each other’s key features, going beyond simple concatenation to model the coordination essential for tone classification.

\begin{figure*}[htbp]
  \centering
  \includegraphics[width=0.9\textwidth]{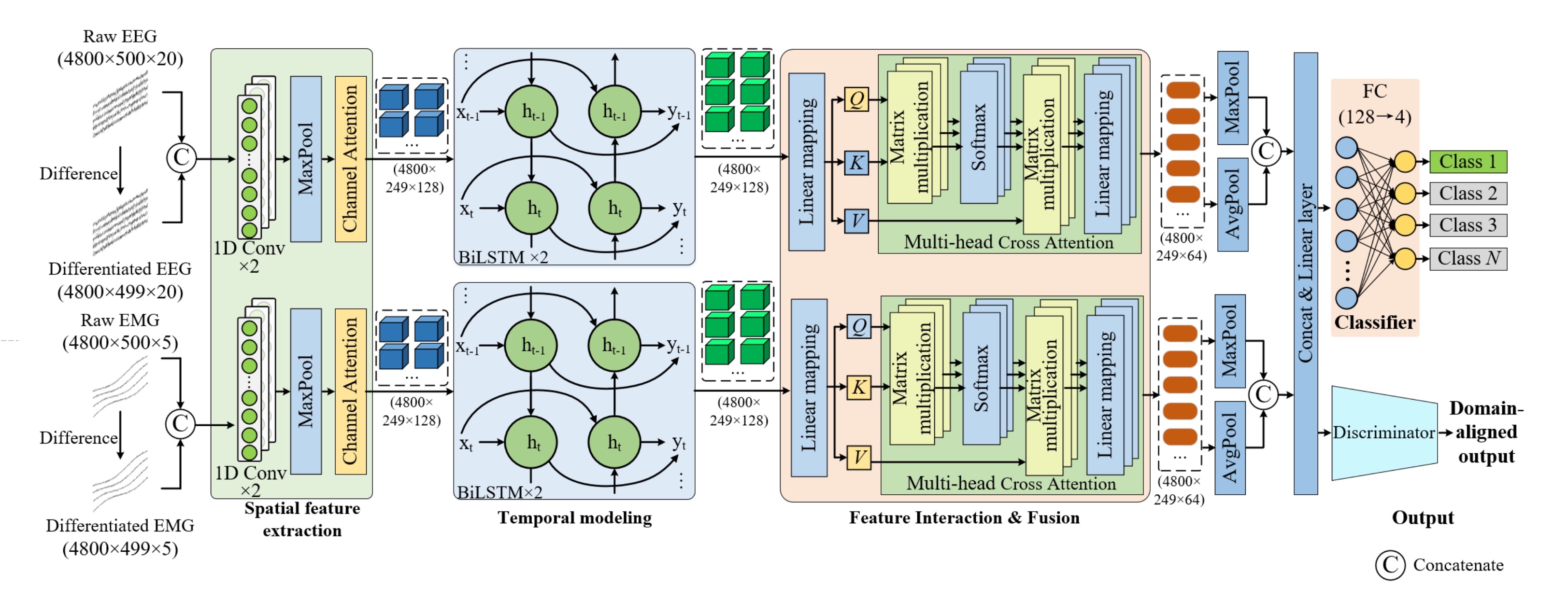}
  \caption{The architecture of our proposed CATNet.}
  \label{fig:wide_figure}
\end{figure*}

\subsection{Cross-Subject Generalization}
Subject variability remains a key challenge for BCI speech systems. EEG and EMG signal patterns differ highly across individuals due to anatomical, physiological, and electrode placement differences, often causing a model to fail on unseen subjects. General domain adaptation frameworks have shown that treating multiple sources as a unified domain can be suboptimal \cite{zhu2019aligning}. For EEG and EMG specific applications, Zhang et al. apply a conditional domain adversarial network to EMG-based silent speech recognition, while Zhao et al. introduce DRDA, a domain discriminator trained adversarially, to encourage the feature extractor to learn domain-invariant representations, thereby improving generalization to unseen subjects or sessions \cite{zhang2023EEGDANN, zhao2021drda}. These approaches underscore the importance of learning features that capture task-relevant information while minimizing subject-specific variance. Inspired by this, our CAT-Net integrates domain-adversarial training into its architecture to explicitly disentangle class-discriminative features from subject-dependent patterns. This enables the model to maintain high performance when deployed to new subjects, particularly enhancing accuracy on abnormal unseen subjects, which is a critical requirement for practical BCI systems.

\section{Methods}

To effectively model the neural-muscular coordination in speech, we propose a model that we call CAT-Net, [C]ross-[A]ttention [T]one Net, a three-stage architecture designed for multimodal EEG-EMG fusion (Fig. 1).  First, spatial-temporal features are extracted independently from EEG and EMG signals. Second, a cross-attention module, inspired by the Transformer architecture \cite{vaswani2017attention}, enables dynamic information exchange between modalities, allowing each to attend to the most informative features of the other. Finally, the fused features serve dual purposes: they are directly fed to a tone classifier to predict the four Mandarin tones, and simultaneously passed through a gradient reversal layer to a domain discriminator that encourages subject-invariant feature learning. In addition, we leverage SHAP (SHapley Additive exPlanations) values to interpret the model's decision process and quantify the relative contributions of EEG and EMG features (see Appendix D).

\subsection{Signal Inputs}
Let $X_i=\{X_i^{e},X_i^{m}\}_{i=1}^{n}$ denote the paired electrophysiological recordings for trial $i$, where the superscripts $e$ and $m$ stand for \textbf{EEG} and \textbf{EMG}, respectively. The labels are $Y_i=\{Y_i^{t},Y_i^{s}\}_{i=1}^{n}$, with $Y_i^{t}\in\{1,2,3,4\}$ encoding the Mandarin tone and $Y_i^{s}$ identifying the speaker. After preprocessing, each trial is represented by the sequence $x=(x_1,\dots,x_T)\in\mathbb{R}^{2C_e+2C_m}$ with $T=500$, where $C_e$ and $C_m$ are the numbers of EEG and EMG channels. Every raw channel is augmented with its \textbf{first-order temporal difference} $\Delta x_t=x_t-x_{t-1}$, a finite-difference operator that accentuates rapid transients, improves stationarity, and has been shown to boost decoding accuracy in both EEG- and EMG-based BCIs \cite{andreou2016temporal, phinyomark2013emg}. Because $\Delta x_t$ is undefined at $t=1$, the effective sequence length becomes $T'=T-1=499$; concatenating the original and differential streams explains the factor 2 in the feature dimension $2C_e+2C_m$.

\subsection{Spatial and Temporal Encoders}
For each of the modalities of EEG and EMG, CAT-Net needs to first encode their spatial and temporal features. For spatial, we use two 1x1 pointwise Conv1D layers. This means we apply a kernel with size of $1 \times 2C_{EEG \text{ or } EMG} \times F$, where $F$ is the feature dimension of size 64 and 128, to each individual timestep $t$.
\begin{equation}
\mathbf{H}_t = \text{ReLU}(X_tW_{conv}) \in \mathbb{R}^{1 \times F}
\end{equation}
Intuitively, this means each column of the kernel weight matrix $W_{conv}$ learns a spatial combination of our EEG or EMG channels into a feature in $F$. A 1D MaxPooling layer with a kernel size of 2 and stride of 2 halves the sequence dimension from 499 to 249, preserving key signal features like neural spikes while reducing computational complexity. To reduce the noise inherent in EEG \& EMG data, we adopt the channel attention from CBAM \cite{woo2018cbam} to re-weight feature importance. We apply a global average pooling to capture general channel scales and global max pooling to capture spikes across our feature representations. $s_a = \text{GlobalAveragePooling}(\mathbf{H}) \in \mathbb{R}^{F}$ and $s_m = \text{GlobalMaxPooling}(\mathbf{H}) \in \mathbb{R}^{F}$. We then pass $s_a$ and $s_m$ to dense layers with sigmoid activation, normalizing the outputs to 0 to 1, $s'_a = \text{sigmoid}(W_2\text{ReLU}(W_1s_a))$. Similarly, $s'_b = \text{sigmoid}(W_2 \text{ReLU}(W_1 s_m))$. To combine these two, we perform a simple sum, $s' = s'_a + s'_b \in \mathbb{R}^F$. The final weights are then broadcasted across the time dimension $T$.
\begin{equation}
\mathbf{\tilde{H}} = s' \odot \mathbf{H} \in \mathbb{R}^{T \times F}
\end{equation}
We then use BiLSTM to model temporal dynamics with the same output of 64 features, as it captures long-range dependencies and bidirectional patterns essential for detecting tonal variations, while being more stable and data-efficient than a full Transformer encoder. Due to having both a forward LSTM and backwards LSTM and concatenating their outputs, the final output is of shape $\mathbf{Z} \in \mathbb{R}^{T \times 2F}$.

\subsection{Cross-Modal Attention Fusion}
For the encodings above, we apply the same operations to both EEG and EMG, each has its unique set of weights. Thus, we have inputs $\mathbf{Z}^{EEG}, \mathbf{Z}^{EMG}$ to our cross attention layer. For each of these input we have three weight matrices $W_Q, W_K \in \mathbb{R}^{2F \times K} ; W_V \in \mathbb{R}^{2F \times V}$ and corresponding outputs $Q^{(e, m)} = \mathbf{Z}^{(e,m)} W_Q^{(e, m)} \in \mathbf{R}^{T \times K}$. We let EEG query the key and value outputs of EMG and vice versa (Algorithm 1, lines 3-4).
\begin{equation}
\mathbf{C}^{(e, m)} = \text{MHA}(Q^{(e, m)}, W^{(m, e)}, K^{(m,e)})
\end{equation}
For each of EEG and EMG, we use a Multi-Head Cross Attention layer with 4 heads and dimension $K = V = 32$. The final output $\mathbf{C}^{(e,m)}$ is of shape $T \times 128$. We apply Layer Normalization to prevent exploding gradients. We then use a global avg pooling and global max pooling to capture activations and spikes across time, $P_{avg} = \text{GlobalAvgPool}(C) \in \mathbb{R}^{1 \times 128}$ and concatenate them, $P = \text{Concat}(P_{avg}, P_{max}) \in \mathbb{R}^{1 \times 256}$. This passes through a simple Dense layer mapping the 256 to 128, then finally is used in our Tone classifier to predict the four Mandarin tones $y \in \{1, 2, 3, 4\}$.

% -------------------------------------------------------
% CAT-Net — forward pass and composite-loss computation
% -------------------------------------------------------
\begin{algorithm}[tb]
\caption{Single training iteration for CAT-Net}
\label{alg:catnet}
\small

\textbf{Input:} EEG mini-batch $\tilde{X}^{e}$, EMG mini-batch $\tilde{X}^{m}$, tone labels $\tilde{Y}^{t}$, subject labels $\tilde{Y}^{s}$\\
\textbf{Parameter:} modality-specific encoders $\mathcal{E}^{e}_{\theta}$ and $\mathcal{E}^{m}_{\theta}$ (same arch, different weights); cross-attention module $\text{XAttn}_{\xi}$; tone classifier $\mathcal{C}_{\phi}$; domain discriminator $\mathcal{D}_{\psi}$; weights $\lambda_{\text{dom}}$; focal hyper-$\gamma$, $\boldsymbol{\alpha}$\\
\textbf{Output:} losses $\mathcal{L}_{\text{focal}}, \mathcal{L}_{\text{dom}}$

\begin{algorithmic}[1]
% ---------- modality encoders ----------------------------------------
\STATE $Z^{(e)} \gets \mathcal{E}^{e}_{\theta}(\tilde{X}^{e})$
\STATE $Z^{(m)} \gets \mathcal{E}^{m}_{\theta}(\tilde{X}^{m})$
% ---------- cross-modal attention ------------------------------------
\STATE $C^{(e)} \gets \text{XAttn}(Q\!=\!Z^{(e)},\;K\!=\!Z^{(m)},\;V\!=\!Z^{(m)})$
\STATE $C^{(m)} \gets \text{XAttn}(Q\!=\!Z^{(m)},\;K\!=\!Z^{(e)},\;V\!=\!Z^{(e)})$
\STATE $Z \gets \text{Fuse}\bigl(C^{(e)},\,C^{(m)}\bigr)$
% ---------- classification & losses ----------------------------------
\STATE $\hat{Y}^{t} \gets \mathcal{C}_{\phi}(Z)$
\STATE $\hat{Y}^{s} \gets \mathcal{D}_{\psi}\!\bigl(\text{GRL}_{\lambda_{\text{dom}}}(Z)\bigr)$
\STATE $\mathcal{L}_{\text{focal}} \gets \text{FocalLoss}(\hat{Y}^{t}, \tilde{Y}^{t})$
\STATE $\mathcal{L}_{\text{dom}} \gets \text{CrossEntropy}(\hat{Y}^{s}, \tilde{Y}^{s})$
\STATE $\displaystyle
       \mathcal{L} = \mathcal{L}_{\text{focal}}
       + \lambda_{\text{dom}}\,\mathcal{L}_{\text{dom}}$
\STATE Back-propagate and update $\theta,\xi,\phi,\psi$ with Adam; update each $c_{k}$ with EMA
\RETURN $\mathcal{L}_{\text{focal}},\;\mathcal{L}_{\text{dom}}$
\end{algorithmic}
\end{algorithm}

\subsection{Domain Discriminator and Loss Functions}
To make CAT-Net maintain accuracy on unseen subjects, we attach a \textbf{domain-discriminator head} to the fused feature $\mathbf{f}\in\mathbb{R}^{128}$ during training. A \textit{gradient-reversal layer} \cite{ganin2016domain} leaves the forward pass unchanged but flips the sign of the gradient, $\mathcal{R}_\lambda(\mathbf{f})=\mathbf{f},\frac{\partial \mathcal{R}_\lambda}{\partial \mathbf{f}}=-\lambda\mathbf{I}$ so that the discriminator learns to predict the subject label $d$ while the backbone is forced toward \textbf{subject-invariant} representations. The network is trained with three losses (Algorithm 1, lines 8-15):
\begin{itemize}
    \item \textbf{Focal loss} $\mathcal{L}_{\text{focal}}$ \cite{lin2017focal} on the tone logits, with $\gamma=2$ and class-balancing vector $\boldsymbol{\alpha}=(0.2,\ 0.3,\ 0.2,\ 0.3)$. Our loss weighting strategy considers tone-level confusion patterns: more confused tones, such as tone 2 and tone 4, are emphasized to enhance discrimination, while less confused tones (e.g., tone 1 and tone 3) are still balanced to prevent bias.
    \item \textbf{Domain Discriminator} $\mathcal{L}_{\text{dom}}$ on the GRL-filtered branch, promoting invariance across the ten subjects.
\end{itemize}
The total objective is a weighted sum which improves leave-one-subject-out accuracy and tightens class clusters.
\begin{equation}
\mathcal{L}
=\mathcal{L}_{\text{focal}}
+0.05\,\mathcal{L}_{\text{dom}}
+(0.2,\ 0.3,\ 0.2,\ 0.3)\,\mathcal{L}_{\text{cent}}
\end{equation}

\section{Experiment}

\subsection{Experiment Setup}

\subsubsection{Dataset}
Ten healthy native Mandarin-speaking adults (aged 24–35 years) participated in this study. To minimize potential confounding factors, all participants were instructed to abstain from consuming coffee, alcohol, or any other substances that might influence their neurophysiological states for 24 hours before the experimental session. Additionally, participants were required to: (1) maintain adequate sleep duration the night before testing; (2) thoroughly cleanse their scalp to ensure optimal electrode contact; and (3) review and sign informed consent documents. The study protocol was approved by our institutional ethics committee and strictly followed the ethical guidelines outlined in the Declaration of Helsinki ~\cite{WMA2013Helsinki}.

\subsubsection{Data Collection}
The experimental paradigm employed four phonologically distinct Mandarin tones, with each tonal category represented by 30 carefully selected Chinese characters that were balanced for lexical frequency and phonological characteristics. Each character were presented four times in a pseudorandomized sequence to prevent habituation effects, with complete counterbalancing of presentation order across all tonal categories (Fig. 2a). During each trial, characters were displayed centrally on the screen for 1.5 seconds, during which participants performed either silent or audible articulation according to the experimental condition, followed by a jittered inter-stimulus interval (1.5-3 seconds) to minimize anticipatory neural responses. The complete stimulus set was presented in both articulation modes, with all participants completing both conditions. Neural activity was recorded using a high-density 64-channel NeuSen W wireless EEG system (Neuracle Technologies, China) synchronized with electromyographic (EMG) recordings from five orofacial muscles (the right buccinator, right cervical trapezius, left buccinator, left cervical trapezius, and mentalis). All physiological signals were acquired at 1000 Hz sampling frequency. The experimental protocol continued until all 480 trials (4 tones × 30 characters × 4 repetitions) were completed in both articulation conditions.
\begin{figure}[t]
    \centering
    \includegraphics[width=0.95\linewidth]{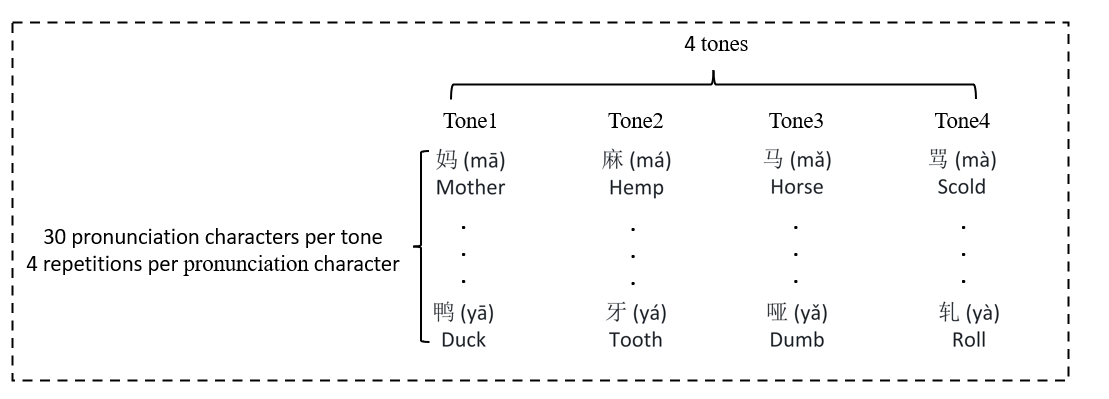}
    \caption*{(a)}
    \vspace{0.5em}
    \includegraphics[width=0.95\linewidth]{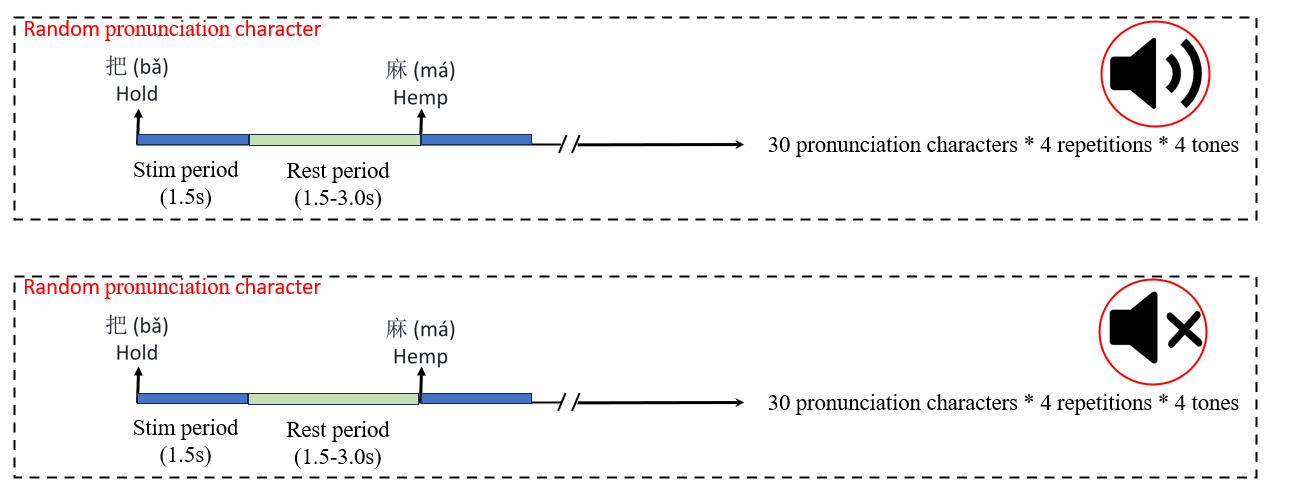}
    \caption*{(b)}
    \caption{Experimental paradigm. (a) Stimulus design showing four tone categories, with thirty distinct pronunciation characters per tone category. Each pronunciation character was presented four times throughout the experiment. (b) Schematic representation of the experimental procedure, illustrating randomized character presentation in both silent and audible modes.}
    \label{fig:catnet}
\end{figure}
\subsubsection{Preprocessing}
For EEG signal preprocessing, the raw signals from valid channels were first downsampled to 500 Hz to improve computational efficiency. A fourth-order Butterworth bandpass filter was applied for frequency band extraction, followed by a 50 Hz notch filter to eliminate power line interference. Common average referencing (CAR) was employed to reduce common-mode noise across channels. Subsequently, independent component analysis (ICA) was performed to identify and remove ocular (EOG) and muscular (EMG) artifacts. Epochs were extracted from (-2 s, 2 s) relative to stimulus onset, with baseline correction applied using the pre-stimulus interval (-2 to 0 s) to eliminate baseline drift. For EMG signals preprocessing, the EMG signals were downsampled to 500 Hz to match the EEG sampling rate. A fourth-order Butterworth bandpass filter was applied, followed by a 50 Hz notch filter to remove power line interference. Common noise was reduced using common average referencing (CAR). Data epochs were extracted from (-2 s, 2 s) relative to stimulus onset, and baseline correction was performed to minimize drift.

\begin{figure}
    \centering
    \includegraphics[width=0.8\linewidth]{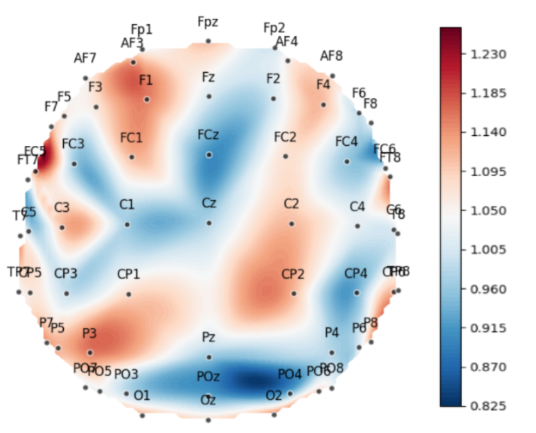}
    \caption{EEG channel weight calculated by Channel-Attention block.}
    \label{fig:enter-label}
\end{figure}

\subsection{Channel Selection}
To identify the most informative EEG channels contributing to tonal classification and improve training efficacy, we incorporated a Channel-Attention block in the early stage of the EEG branch. Inspired by the channel attention mechanism in CBAM~\cite{woo2018cbam}, we apply a global average and max pooling–based gating module to adaptively re-weight each EEG channel. These weights reflect the relative importance assigned to each channel during the classification process and can be aggregated across samples to identify the most informative electrodes. After training, we visualized the normalized attention weights on a 2D topographic scalp map (Fig. 3). As shown, channels located in the frontal, central, and parietal regions exhibit stronger attention responses, suggesting greater contributions to tone classification. This finding is consistent with prior neurophysiological studies that have highlighted the involvement of the prefrontal and parietal cortices in tone and syllable perception~\cite{Ni2023TemporalFineStructure}, while our results further emphasize the potential role of central regions.
\begin{table}[H]
\centering
\resizebox{\linewidth}{!}{
\begin{tabular}{lcccccc}
\toprule
\textbf{Channel} & \textbf{Tone 1} & \textbf{Tone 2} & \textbf{Tone 3} & \textbf{Tone 4} & \textbf{Average} & \textbf{Kappa} \\
\midrule
5    & 99.42 & 79.60 & 85.17 & 82.36 & 86.92 & 0.8249 \\
10   & 98.68 & 80.80 & 86.24 & 81.29 & 87.02 & 0.8251 \\
20   & 98.67 & 83.64 & 87.29 & 83.10 & 88.08 & 0.8415 \\
Full & 99.12 & 83.99 & 88.25 & 84.67 & 88.45 & 0.8496 \\
\bottomrule
\end{tabular}
}
\caption{Performance comparisons under different channel numbers(\%). We list the ranked weighted channels in the Appendix B. Average denotes the average accuracy over K-fold cross-validation runs.}
\label{tab:channel_tone_kappa}
\end{table}

\begin{table*}[htbp]
\centering
\resizebox{0.9\linewidth}{!}{  
\begin{tabular}{lccccccccccc}
\toprule
\textbf{Method} & 
\multicolumn{2}{c}{\textbf{Tone 1}} & 
\multicolumn{2}{c}{\textbf{Tone 2}} & 
\multicolumn{2}{c}{\textbf{Tone 3}} & 
\multicolumn{2}{c}{\textbf{Tone 4}} & 
\multicolumn{2}{c}{\textbf{Macro Avg}} \\
\cmidrule(r){2-3} \cmidrule(r){4-5} \cmidrule(r){6-7} \cmidrule(r){8-9} \cmidrule(r){10-11}
& F & R & F & R & F & R & F & R & F & R \\
\midrule
EEGNet & 98.27 & 98.92 & 80.14 & 77.33 & 85.66 & 86.58 & 76.95 & 78.33 & 85.29 & 85.26 \\
DeepConvNet & 93.52 & 85.17 & \textbf{83.91} & \textbf{85.18} & 87.53& \textbf{91.17} & 80.93 & 80.08 & 86.56 & 86.64 \\
GAT & 97.25 & 97.88 & 75.21 & 67.15 & 81.94 & 89.93 & 77.87 & 78.55 & 83.62 & 83.60 \\
CATNet & \textbf{99.54} & \textbf{99.83} & 81.45 & 80.36 &\textbf{ 87.91} & 87.58 & \textbf{82.54} & \textbf{84.33} & \textbf{88.08} & \textbf{88.06 }\\
\bottomrule
\end{tabular}
}
\caption{Comparison results on each tone's F-1 score and recall with three baseline methods with ranked highest average precisions. “F” and “R” represent “F1-score” and “Recall” respectively. The results are in the silent speech condition. All numbers are in \%.}
\label{tab:silent_tone_results}
\end{table*}

Based on this channel attention selection, we conduct experiments on the balance of channels and training efficacy, finding the least EEG channel to train without significantly lowering our prediction accuracy. In Table 1, we select the top five, ten, and twenty channels with the highest weights calculated by channel attention and compare them with the full channels’ accuracy to investigate the balance between model efficacy and accuracy. Based on the results from Table 1, we observe that training with 20 channels reaches a balance between trimming less informative channels and high accuracy. Thus, subsequent experiments will primarily focus on models utilizing a reduced set of 20 EEG channels in the frontal, central, and parietal brain regions.

\subsection{Baseline Comparisons}

\subsubsection{K-Fold Cross Validation}
 To validate the accuracy of our proposed EEG-EMG fusion model, we compared our algorithms in both silent and audio conditions against eight main state-of-the-art EEG decoding algorithms, including End-to-End CNN~\cite{Wang2023EndToEndTone}, VLAAI~\cite{Accou2023VLAAI}, EEGNet~\cite{lawhern2018eegnet}, DeepConvNet~\cite{Schirrmeister2017DeepConvNet}, FBCSP+SVM~\cite{Ang2008FBCSP,Cortes1995SVM}, GAT~\cite{song2023global}, DRDA~\cite{zhao2021drda}, and EEG-Transformer~\cite{eegtrans}. We reproduced these algorithms on our datasets and compared their performance with our proposed CATNet model. For algorithms that are not utilizing EMG signals, we concatenate EEG and EMG signals as input to the model. We use average precision and kappa values as metrics. To mitigate overfitting and ensure robust performance evaluation, we employed 5-fold cross-validation during model training and testing. For brevity, we present the results for the audio speech condition in Appendix C. The subsequent analysis centers on model performance under the silent speech condition.

\begin{table}[t]
\centering
\resizebox{\linewidth}{!}{  
\begin{tabular}{lcccccc}
\toprule
\textbf{Method} & \textbf{Tone1} & \textbf{Tone2} & \textbf{Tone3} & \textbf{Tone4} & \textbf{Average} & \textbf{Kappa} \\
\midrule
ETE-CNN & 71.15 & 69.16 & 46.59 & 48.18 & 57.33 & 0.3725\\
VLAAI & 94.21 & 39.34 & 69.84 & 49.98 & 61.12 & 0.5515\\
EEGNet & 98.10 & 83.15 & 84.75 & 75.62 & 85.29 & 0.8037 \\
DeepConvNet & \underline{99.57} & 82.69 & 83.75 & \underline{81.79} & \underline{86.56}& \underline{0.8125} \\
FBCSP+SVM & 87.30 & 41.01 & 39.04 & 37.24 & 51.65& 0.3553 \\
GAT &97.72&73.18&\textbf{88.56}&74.33&83.62&0.7888\\
DRDA & \textbf{99.71 }& \underline{83.27} & 80.22 & 61.01 & 81.23 & 0.7802 \\
EEG-Transformer & 92.25 & 79.86 & 82.63 & 69.94 & 81.10 & 0.7477 \\
CATNet & 98.67 & \textbf{83.64} & \underline{87.27} &\textbf{ 83.10} & \textbf{88.08} & \textbf{0.8415}\\

\bottomrule
\end{tabular}
}
\caption{Tone precision with baseline comparisons in the silent speech condition. ETE-CNN represents End-to-End CNN. Average Precision and Kappa metrics are employed in the table. The best results in each column are highlighted in bold, and the second-best are underlined. The results are under 20 EEG channels and the silent speech condition.}
\end{table}

\begin{table*}[ht]
\centering
\resizebox{0.9\textwidth}{!}{
\begin{tabular}{lcccccccccccc}
\toprule
\textbf{Method} & S1 & S2 & S3 & S4 & S5 & S6 & S7 & S8 & S9 & S10 & Avg \\
\midrule
ETE-CNN & 38.12 & 30.00 & 71.67 & 36.67 & 70.21 & 83.13 & 49.38 & 70.00 & 25.42 & 54.79 & 52.94 \\
VLAAI & 46.53 & 50.26 & 56.72 & 60.46 & 65.43 & 63.56 & 62.96 & 55.87 & 62.38 & 60.45 & 58.47 \\
EEGNet      & 25.83 & 79.17 & 75.00 & 78.96 & 79.37 & 81.25 & 92.92 & 82.92 & 95.42 & 87.08 & 77.79 \\
DeepConvNet & 41.67 & 73.75 & 76.46 & 82.50 & 84.58 & 86.88 & 90.42 & 85.83 & \textbf{95.63} & 87.08 & 80.48 \\
FBCSP+SVM & 38.37 & 43.36 & 48.31 & 50.24 & 52.06 & 57.91 & 57.49 & 42.93 & 55.83 & 51.02 & 49.75 \\
GAT         & 43.75 & 78.75 & 71.04 & 79.79 & 82.29 & 90.42 & 89.58 & 87.29 & 92.08 & 90.42 & 80.54 \\
DRDA        & 41.15 & 79.24 & 71.36 & 75.32 & 82.28 & 89.17 & 61.93 & 84.24 & 92.54 & 82.48 & 75.37 \\
EEG-Transformer  & 27.50 & 53.33 & 63.33 & 76.25 & 66.25 & 63.96 & 81.87 & 82.29 & 86.04 & 84.38 & 68.52 \\
\textbf{CATNet} & \textbf{53.33} & \textbf{79.58} & \textbf{82.71} &\textbf{ 85.00} & \textbf{87.08} & \textbf{91.67} & \textbf{95.00} & \textbf{89.17} & 94.79 & \textbf{92.71} & \textbf{85.10}\\
\bottomrule
\end{tabular}
}
\caption{Cross-subject classification performance on ten subjects. (Silent Speech Condition).}

\label{tab:cross_subject_silent}
\end{table*}

\begin{table*}[ht]
\centering
\resizebox{0.9\textwidth}{!}{
\begin{tabular}{lcccccccccccc}
\toprule
\textbf{Method} & S1 & S2 & S3 & S4 & S5 & S6 & S7 & S8 & S9 & S10 & Avg \\
\midrule
\textit{CATNet\textsubscript{w/o DD}} & 48.96 & 80.83 & 82.92 & 84.17 & 87.29 & 92.50 & 93.54 & 89.38 & 94.79 & 92.50 & 84.69 \\
\textit{CATNet} & \textbf{53.33} & 79.58 & 82.71 & \textbf{85.00} & 87.08 & 91.67 &\textbf{ 95.00} & 89.17 & 94.79 & \textbf{92.71} & 85.10 \\
\bottomrule
\end{tabular}
}
\caption{Cross-subject ablation results. "DD": Domain Discriminator.}

\label{tab:cross_subject_silent}
\end{table*}
As shown in Table 3, our proposed method achieves the highest average accuracy of 88.08\% and kappa value of 0.8415 in the silent speech condition. For audio speech, we achieve 87.83\% accuracy with kappa value of 0.8377. Both results outperform all baseline models. In particular, CATNet demonstrates superior performance on the challenging Tone 2 and Tone 4, which are traditionally difficult to distinguish due to their subtle frequency contours.

Table 2 presents F1 scores and recall for comprehensive evaluation. The three baselines are selected based on their high average precision in Table 3. CATNet achieves the highest F1 and recall for Tone 1 (F=99.54, R=99.83) and Tone 4 (F=82.54, R=84.33). Compared to DeepConvNet, CATNet improves Tone 4 recall (84.33 vs. 80.08) and achieves stronger macro-averaged performance (F1=88.08, R=88.06). While DeepConvNet shows slightly higher Tone 2 performance, CATNet achieves better precision (83.64 vs. 82.69), indicating fewer false alarms and more conservative predictions beneficial for high-stakes scenarios. The consistent performance across tones confirms that CATNet provides superior discriminative and generalizable representations for tone-level silent speech decoding.

\subsubsection{Cross-Subject Calibration}
In general, EEG and EMG signals in BCI applications exhibit strong cross-subject variability, which poses significant challenges to model generalization on unseen individuals. To address this limitation, we conduct cross-subject evaluations to assess the transferability of our proposed model across different participants. We used a standard LOSO (Leave-One-Subject-Out) training strategy. We withheld data from a single subject as the test set and used the nine remaining subjects’ data for training. Table 4 shows the results in silent speech conditions, and audio speech baseline comparisons are in Appendix C. 

The results demonstrate our model's strong generalization ability under cross-subject evaluation, with accuracy dropping slightly \textbf{2.98\%} from 88.08\% to 85.10\%. In contrast, previously strong-performing baselines exhibit greater deviation in accuracy, indicating a weakness in the model's performance on unseen subjects. For example, EEGNet and DeepConvNet experience accuracy drops of \textbf{7.5\%} and \textbf{6.08\%}, respectively. Notably, the accuracy for each subject, except S9 (0.84\% lower), outperforms the second-best baseline by an average of \textbf{2.88\%} across the remaining subjects, further confirming our significant superiority across all subjects. Similarly, our model accuracy dropped from 87.83\% to 83.27\% in the audio speech condition, while other baseline models also suffer from greater deviation. Furthermore, our model shows strong improvement in abnormal subjects such as S1(\textbf{6.8\%} higher than the second-best baseline), S3(\textbf{6.25\%} higher), and S4(\textbf{2.5\%} higher), where baseline models struggle to generalize the learned features to abnormal individuals. Therefore, it can be shown that our proposed method is capable of handling individual differences and generating stable accuracy even with abnormal subjects' data.

\begin{table}[ht]
\centering
\scriptsize
\setlength{\tabcolsep}{4pt}  
\begin{tabular}{lccccccccc}
\toprule
\textbf{Method} & \textbf{CT} &\textbf{TF} & \textbf{FF}  & \textbf{DD}  & \textbf{P} & \textbf{R} & \textbf{F1} \\
\midrule
\textit{CATNet\textsubscript{w/o ct}}    &  &\ding{51}  &\ding{51}   &\ding{51}  &   77.63 & 77.62& 77.00 \\
\textit{CATNet\textsubscript{w/o BiLSTM}}    & \ding{51} &   & \ding{51}   &\ding{51}  & 78.42 & 78.42& 78.41 \\
\textit{CATNet\textsubscript{w/o fusionEMG}}    &  & \ding{51} &     & \ding{51} & 76.60  & 76.60  & 76.50  \\
\textit{CATNet\textsubscript{w/o fusionEEG}}    &  & \ding{51} &     & \ding{51} & 76.46   &76.46  & 76.35  \\

\textbf{CATNet}   & \ding{51} &\ding{51} & \ding{51}  & \ding{51}   & \textbf{88.08} & \textbf{88.08} & \textbf{88.06} \\
\bottomrule
\end{tabular}
\caption{Ablation studies in 5-fold training scenario. Legend: “CT”: Cross-Attention, “TF”: Temporal Feature with BiLSTM layer, “FF”: Fusing Feature, "DD": Domain Discriminator. “w/o.” is short for “without”.
 }
\label{tab:catnet_ablation}
\end{table}

\subsection{Ablation Study}
We conducted ablation experiments to assess the contribution of each module in CATNet. The results of 5-fold validation and cross-subject generalization are presented in Tables 5 and 6, respectively. Our analysis focuses on two core aspects of CATNet: (1) the role of cross-attention in multimodal fusion, and (2) the impact of the domain discriminator (DD) on subject-level generalization. 

\subsubsection{Cross-Attention Mechanism}
Cross-attention allows the model to selectively focus on the relevant features between EEG and EMG. To assess its importance, we progressively removed the cross-attention module and its sub-components. As shown in Table 6, removing the entire cross-attention module leads to a substantial drop in precision (from 88.08\% to 77.63\%). Furthermore, when we isolate the effect of each directional attention (i.e., attention from EEG to EMG and vice versa), we observe comparable degradation, with both variants yielding precision scores around 76–77\%. These results confirm that cross-attention is indispensable for capturing inter-modal dependencies to boost recognition accuracy.

\subsubsection{Generalization Across Subjects}
To investigate the contribution of the domain discriminator (DD), we compare CATNet with and without the DD module under the cross-subject setting, as presented in Table 5. Overall, the absence of DD leads to a slight drop in average accuracy (from 85.10\% to 84.69\%), indicating that the model still maintains robust generalization capability even without domain supervision. However, substantial performance gains are observed on individual subjects with the DD module. In particular, for subject S1, accuracy improves significantly from 48.96\% to 53.33\%, highlighting DD’s effectiveness in handling severe subject-specific variations. Since our dataset contains a relatively balanced distribution across subjects and employs a minimal-channel configuration that limits overfitting, the marginal domain gaps are already well-handled by the core network. Therefore, DD may yield greater benefits in low-resource scenarios or in datasets with greater heterogeneity across subjects or recording conditions.

\section{Conclusion}
In the paper, we propose the CATNet method that utilizes a deliberated mechanism, demonstrating the efficacy of a multimodal EEG-EMG fusion framework for tonal speech decoding, achieving high accuracy in both audible and silent speech conditions with minimal-channel configurations. Notably, CATNet maintains strong generalization ability in cross-subject evaluations—a challenging yet crucial setting for real-world BCI applications. These results establish CATNet as a solid baseline for tone-level bio-signal decoding and highlight its potential for low-resource, multimodal speech interfaces. In future work, we aim to generalize our framework to accommodate additional input modalities beyond EEG and EMG, and apply the framework to broader classification tasks. We also aim to validate the generalizability of our architecture across more diverse datasets, further exploring its scalability and potential for real-world deployment in low-resource, multi-modal BCI systems.

\section{Acknowledgments}
This work was supported by Shanghai Yang Fan Foundation (No. 24YF2730700).

\bibliography{aaai2026}

\vspace{5cm}
\section{Appendix}
\subsection{A. Hyperparameter Settings }
\renewcommand{\thetable}{A\arabic{table}}  % 设置为 Table B1、B2…
\setcounter{table}{0}   
\begin{table}[ht]
\centering
\small  % 字体稍微缩小，避免溢出
\begin{tabular}{p{3.4cm}|p{4.0cm}}  % 控制列宽避免超出双栏
\toprule
\textbf{Hyperparameter} & \textbf{Value} \\
\midrule
Optimizer & Adam \\
Learning Rate & 1e-3 \\
Loss Function & Focal Loss ($\gamma$=2.0, $\alpha$=(0.2, 0.3, 0.2, 0.3)) \\
Loss Weights & class: 1.0,\newline domain: 0.05 \\
Dropout Rate & 0.4 \\
Batch Size & 64 \\
Epochs & 50 \\
Cross Validation & 5-Fold \\
LSTM Hidden Units & 128 \\
Early Stopping Patience & 10 \\
LR Scheduler & ReduceLROnPlateau\newline (factor=0.5,\newline patience=5,\newline min\_lr=1e-5) \\
EEG/EMG Channels Used & 20 and 5 \\
\bottomrule
\end{tabular}
\caption{Hyperparameter settings used for all experiments.}
\label{tab:hyperparams}
\end{table}

\subsection{B. Experiment Details}
\renewcommand{\thefigure}{B\arabic{figure}}
\setcounter{figure}{0}
\renewcommand{\thetable}{B\arabic{table}}  % 设置为 Table B1、B2…
\setcounter{table}{0}   
\subsection{EEG and EMG Channel Placements}

We collected EEG and EMG data from participants, and the corresponding electrode placement is illustrated in the figure below.
\begin{figure}[H]
    \centering
    \includegraphics[width=0.99\linewidth]{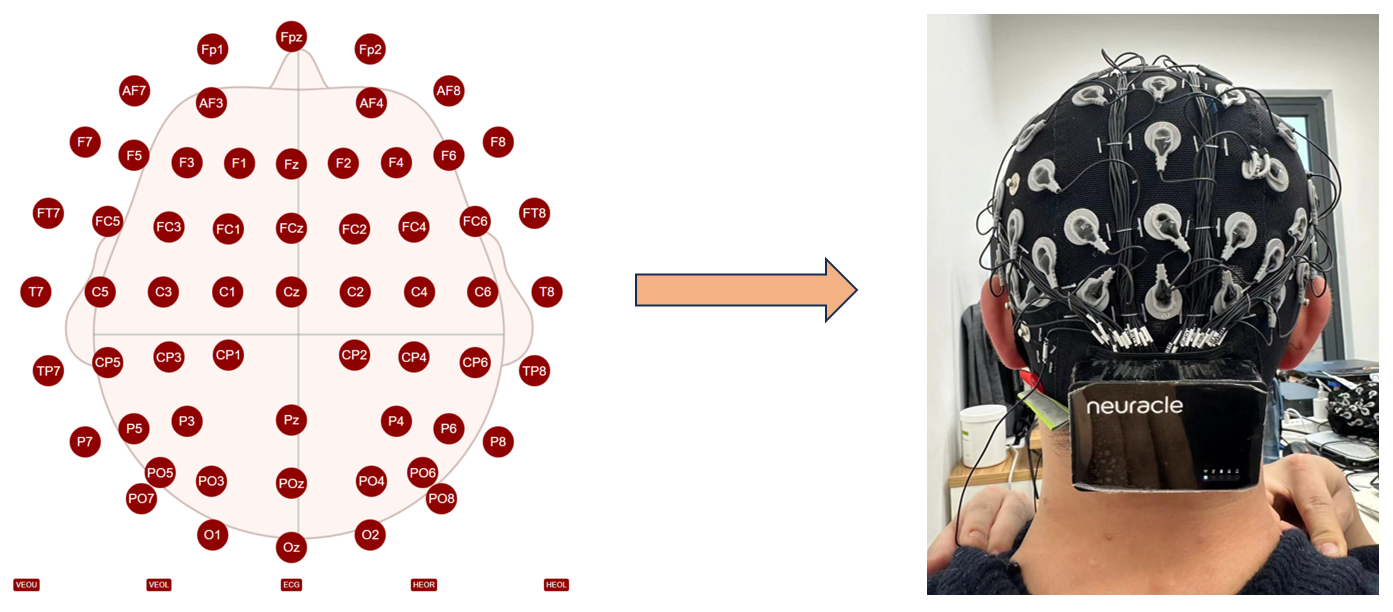}
    \caption{Spatial distribution of EEG electrode placement across the scalp.}
    \label{fig:enter-label}
\end{figure}

\begin{figure}[H]
    \centering
    \includegraphics[width=0.99\linewidth]{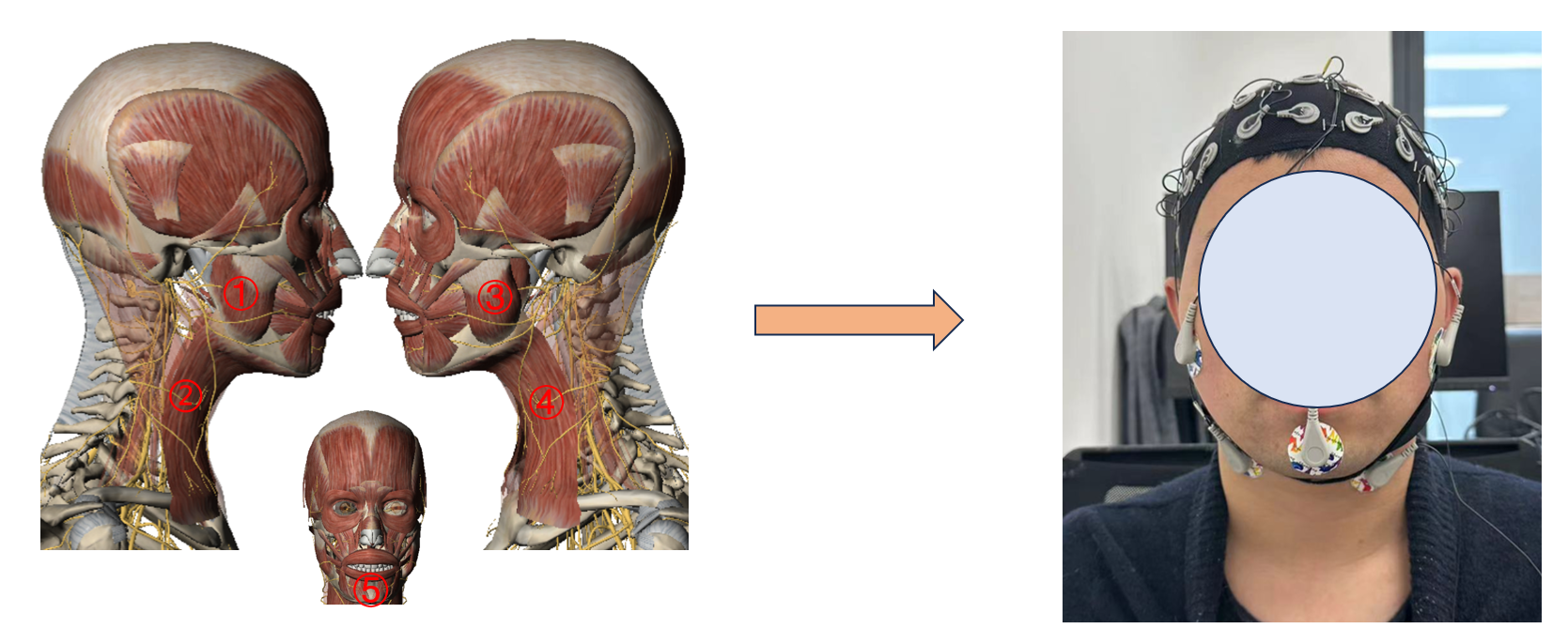}
    \caption{Locations of EMG sensor placement for muscle activity recording. (1. Right Buccinator. 2. Right Cervical Trapezius. 3. Left Buccinator. 4. Left Cervical Trapezius. 5. Mentalis.)}
    \label{fig:enter-label}
\end{figure}

\subsection{Visualization of Temporal Features}
To better understand how to structure the model, we extract and visualize the temporal features of EEG and EMG. For brevity, we present one EEG channel and one EMG channel and their temporal features under both audio and silent speech conditions. From the graph, it can be observed that Tone 2 and Tone 4 exhibit similar EEG and EMG temporal patterns, which explains why our model and all baseline models struggle to differentiate these two tones. 
\begin{figure}[H]
    \centering
    \includegraphics[width=0.8\linewidth]{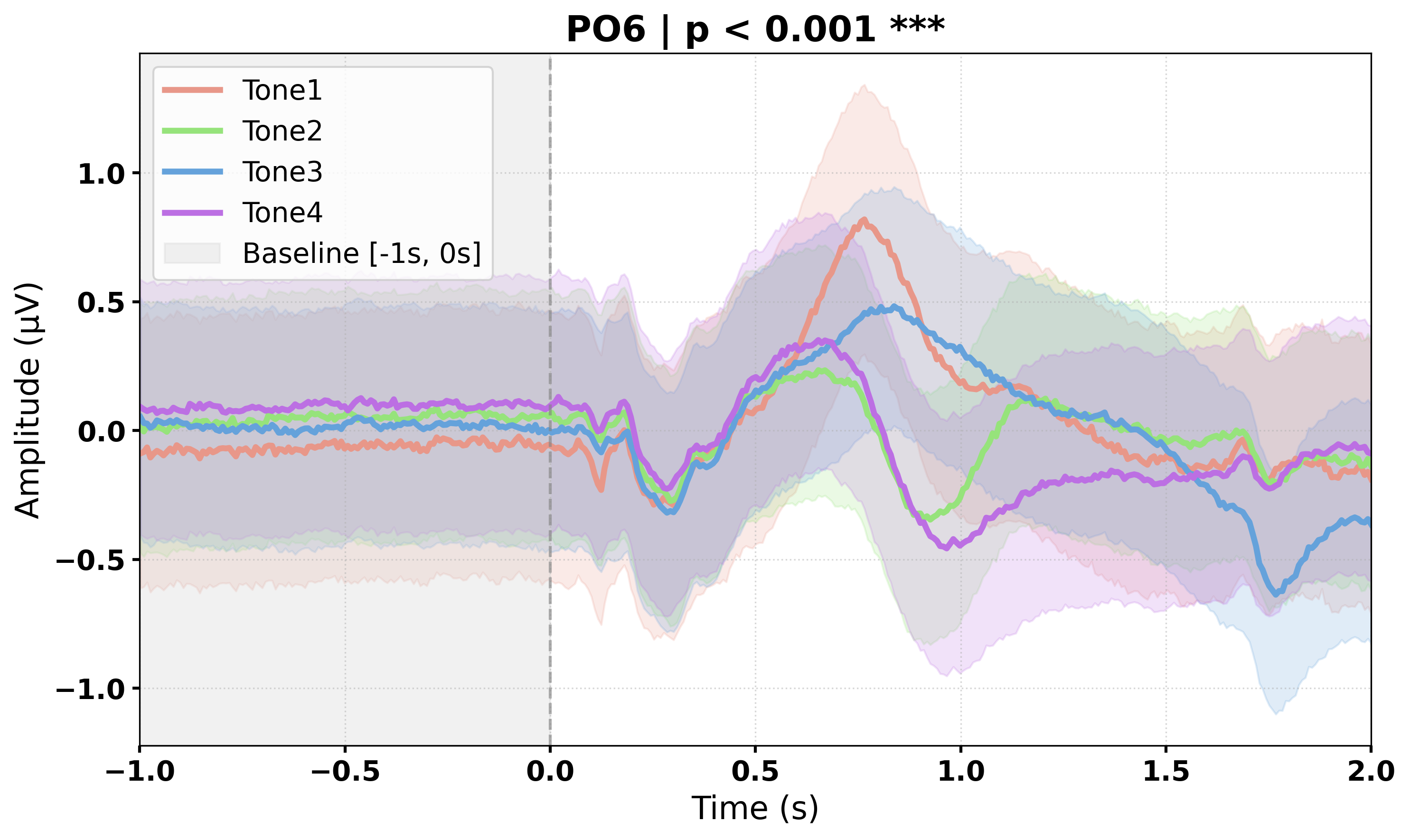}
    \caption{EEG channel PO6 temporal features in audio speech condition.}
    \label{fig:enter-label}
\end{figure}
\vspace{-5mm}
\begin{figure}[H]
    \centering
    \includegraphics[width=0.8\linewidth]{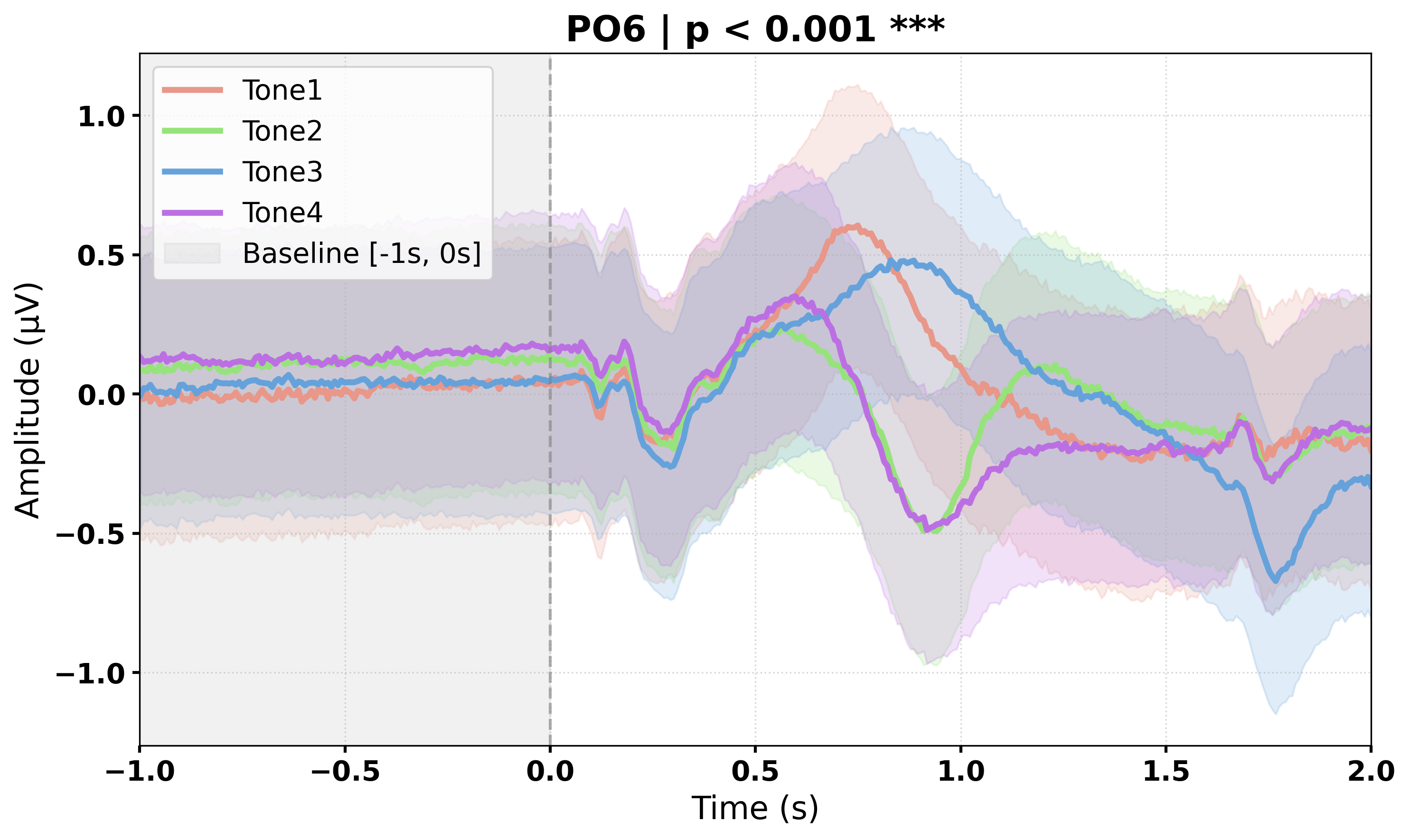}
    \caption{EEG channel PO6 temporal features in silent speech condition.}
    \label{fig:enter-label}
\end{figure}
\vspace{-5mm}
\begin{figure}[H]
    \centering
    \includegraphics[width=0.8\linewidth]{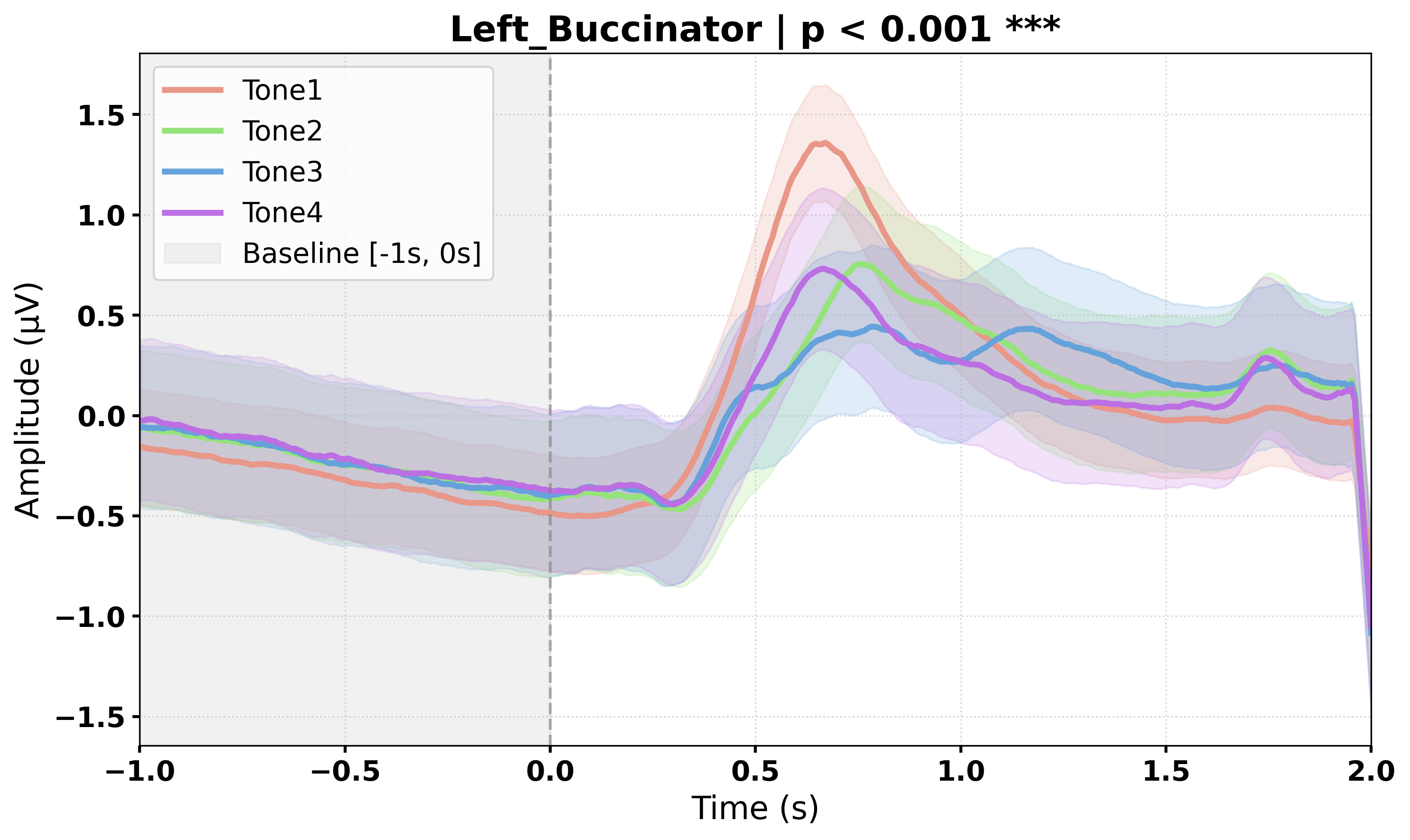}
    \caption{EMG channel Left Buccinator temporal features in audio speech condition.}
    \label{fig:enter-label}
\end{figure}
\vspace{-5mm}
\begin{figure}[H]
    \centering
    \includegraphics[width=0.8\linewidth]{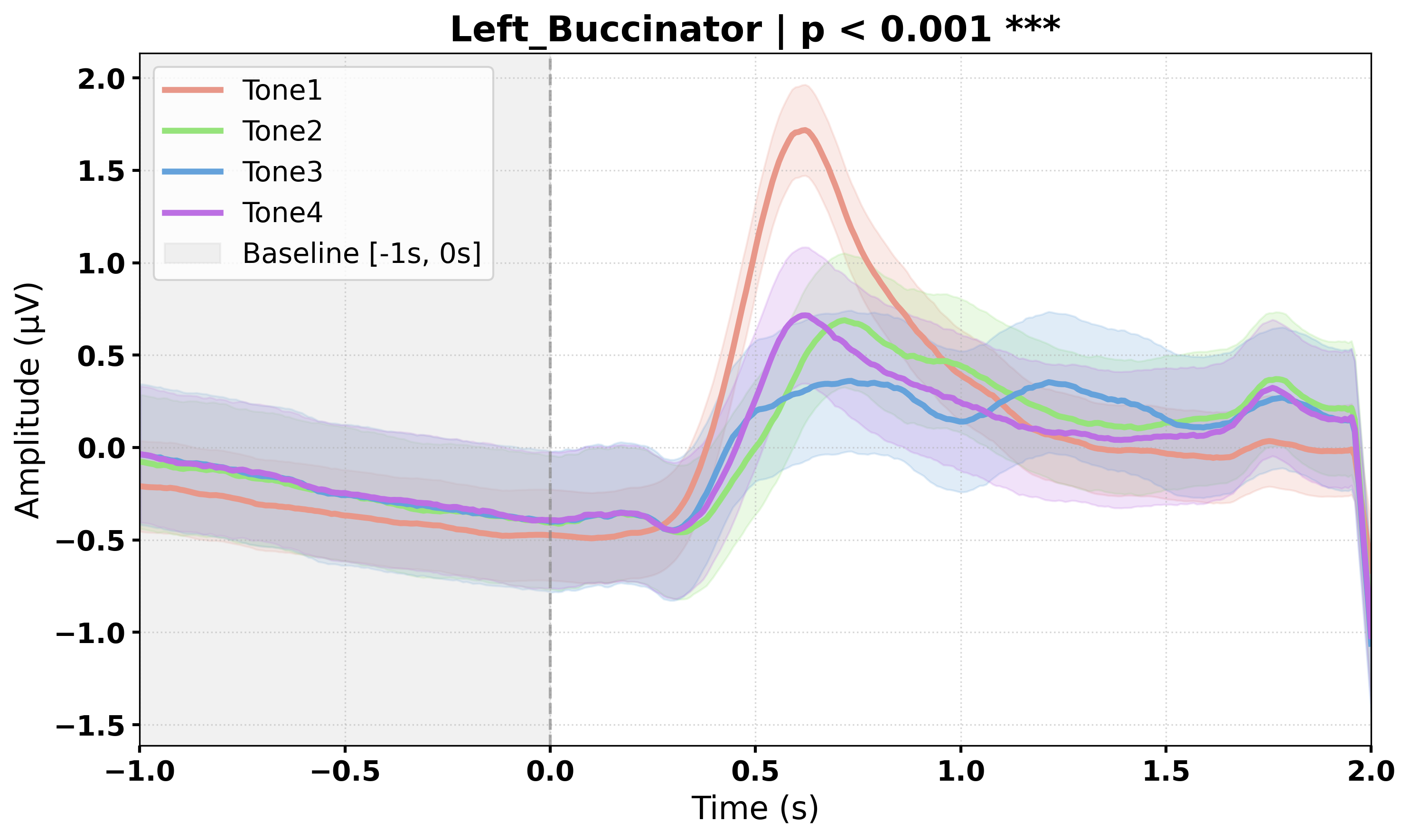}
    \caption{EMG channel Left Buccinator temporal features in silent speech condition.}
    \label{fig:enter-label}
\end{figure}

\subsection{Channel Weights Based On Channel Attention}
\begin{table}[H]
\centering
\label{tab:eeg-importance}
\resizebox{\linewidth}{!}{%
\begin{tabular}{ll|ll|ll|ll}
\toprule
Channel & Weight & Channel & Weight & Channel & Weight & Channel & Weight \\
\midrule
AF7 & 1.1566 & CP4 & 1.0700 & FC5 & 1.0229 & P3 & 0.9864 \\
P03 & 1.1390 & FT8 & 1.0648 & T8 & 1.0223 & FC2 & 0.9815 \\
P06 & 1.1390 & CP3 & 1.0555 & P4 & 1.0199 & FC3 & 0.9814 \\
F7 & 1.1242 & P07 & 1.0532 & POz & 1.0094 & AF4 & 0.9760 \\
FC6 & 1.1193 & C3 & 1.0531 & P7 & 1.0077 & F2& 0.9607 \\
Fp1 & 1.1134 & P05 & 1.0514 & TP8 & 1.0077 & PO8 & 0.9602 \\
CP6 & 1.1100 & F6 & 1.0487 & C2 & 1.0058& Fz & 0.9541 \\
FC4 & 1.1097 & P8 & 1.0430 & CP1 & 1.0056 & Oz & 0.9507 \\
F1 & 1.0793 & C4 & 1.0401 & FCz & 1.0022 & CP2 & 0.9477 \\
PO4 & 1.0772 & AF8 & 1.0343 & Fp2 & 0.9944 & TP2 & 0.9474 \\
O1 & 1.0745 & AF3 & 1.0329 & T7 & 0.9943 & F8 & 0.9404 \\
O2 & 1.0724 & Cz & 1.0303 & Pz & 0.9910 &  &  \\
FT7 & 1.0718 & P6 & 1.0251 & F4 & 0.9878 &  &  \\
\bottomrule
\end{tabular}}
\caption{EEG channel importance weights sorted in descending order. The table is viewed from left to right.}
\end{table}

\subsection{C. Baseline Comparison in Audio Speech Condition}
\renewcommand{\thetable}{C\arabic{table}}  % 设置为 Table C1、C2…
\setcounter{table}{0}   
\subsection{Baseline Models}
\textbf{End-to-End CNN}~\cite{Wang2023EndToEndTone} A convolutional neural network that directly maps raw EEG signals to class labels through stacked 1D convolutional layers and global average pooling.

\textbf{VLAAI}~\cite{Accou2023VLAAI} A modular architecture composed of repeated blocks integrating convolutional layers, layer normalization, and GRUs, designed for decoding EEG sequences.

\textbf{EEGNet}~\cite{lawhern2018eegnet} A compact convolutional neural network that combines temporal convolution, depthwise spatial filtering, and separable convolutions for efficient EEG feature extraction.

\textbf{DeepConvNet}~\cite{Schirrmeister2017DeepConvNet} A deep network with four convolutional-max pooling blocks designed to capture low- to high-level EEG signal features progressively.

\textbf{FBCSP+SVM}~\cite{Ang2008FBCSP,Cortes1995SVM} A traditional pipeline using filter bank common spatial patterns (FBCSP) to extract band-power features followed by a support vector machine classifier.

\textbf{GAT}~\cite{song2023global} A transformer-based network utilizing multi-head self-attention and cross-channel temporal fusiony.

\begin{table*}[htbp]
\centering
\resizebox{\linewidth}{!}{  
\begin{tabular}{lccccccccccc}
\toprule
\textbf{Method} & 
\multicolumn{2}{c}{\textbf{Tone 1}} & 
\multicolumn{2}{c}{\textbf{Tone 2}} & 
\multicolumn{2}{c}{\textbf{Tone 3}} & 
\multicolumn{2}{c}{\textbf{Tone 4}} & 
\multicolumn{2}{c}{\textbf{Macro Avg}} \\
\cmidrule(r){2-3} \cmidrule(r){4-5} \cmidrule(r){6-7} \cmidrule(r){8-9} \cmidrule(r){10-11}
& F & R & F & R & F & R & F & R & F & R \\
\midrule
EEGNet & 98.59 & 99.08 & 75.56 & 75.25 & 86.46 & 85.92 & 74.51 & 74.92 & 83.78 & 83.79 \\
DeepConvNet & 96.42 & \textbf{99.92} & 79.48 & 76.17 & 87.93 & 87.08 & 80.30 & 81.50 & 86.03 & 86.17 \\
GAT & 97.51 & 98.00 & 68.72 & 62.04 & 76.92 & 83.33 & 73.58 & 75.25 & 79.54 & 79.57 \\
CATNet & \textbf{99.50} & 99.67 & \textbf{81.03} & \textbf{80.08} & \textbf{88.76} & \textbf{89.17} & \textbf{81.97} & \textbf{82.42} & \textbf{87.82} &\textbf{ 87.83} \\
\bottomrule
\end{tabular}
}
\caption{Comparison results on each tone's f-1 score and recall with three baseline methods with ranked highest average precisions. “F” and “R” represent “F1-score” and “Recall” respectively. The results are in audio speech condition. All number are in \%.}
\label{tab:silent_tone_results}
\end{table*}

\begin{table*}[ht]
\centering
\resizebox{\textwidth}{!}{
\begin{tabular}{lcccccccccccc}
\toprule
\textbf{Method} & S1 & S2 & S3 & S4 & S5 & S6 & S7 & S8 & S9 & S10 & Avg \\
\midrule
ETE-CNN & 25.00 & 50.62 & 48.54 & 59.38 & 25.00 & 46.80 & 25.00 & 25.00 & 49.58 & 69.37 & 42.44 \\
VLAAI & 51.25 & 55.23 & 55.67 & 54.80 & 61.73 & 62.11 & 64.23 & 72.14 & 72.32 & 64.22 & 61.34\\
EEGNet      & 55.00 & 68.75 & 67.92 & 70.21 & 80.83 & 79.17 & 89.17 & 83.54 & 94.79 & 81.46 & 77.08 \\
DeepConvNet & 49.79 & 58.33 & 72.29 & 80.63 & 83.75 & 80.83 & 85.21 & 79.79 & 87.71 & 85.21 & 76.35 \\
FBCSP+SVM & 33.95 & 41.04 & 46.87 & 43.33 & 47.50 & 54.16 & 53.95 & 48.95 & 56.66 & 53.95 & 48.04\\
GAT         & 51.67 & 64.79 & 67.29 & 71.88 & 85.00 & 85.21 & 85.00 & 84.17 & 86.88 & \textbf{90.00} & 77.18\\
DRDA        & 48.71 & 56.00 & 72.94 & 76.12 & 78.14 & 76.09 & 82.94 & 79.23 & 87.71 & 84.23 & 74.24\\
EEG-Transformer    & 49.79 & 47.92 & 57.29 & 74.79 & 76.04 & 80.21 & 79.58 & 67.08 & 55.21 & 58.54 & 64.65\\
\textbf{CATNet} & \textbf{56.04} &\textbf{ 72.08} & \textbf{75.42} & \textbf{80.63} & \textbf{92.08} & \textbf{90.00} & \textbf{91.87} & \textbf{90.62} & \textbf{94.79} & 89.17 & \textbf{83.27}\\
\bottomrule
\end{tabular}
}
\caption{Cross-subject classification performance on ten subjects. }

\label{tab:cross_subject_silent}
\end{table*}

\textbf{DRDA}~\cite{zhao2021drda} A domain-adversarial architecture combining a shared feature extractor with a domain discriminator and gradient reversal layer.

\textbf{EEG-Transformer}~\cite{eegtrans} A transformer based network with self-attention applied to EEG signals, enabling efficient modeling of long-range temporal and spatial dependencies for brain-computer interface tasks.

\subsection{K-Fold Results}
In this subsection, we run all the baselines and our proposed CATNet under the audio speech condition. The table is shown below in Table C3. As shown in Table C3, our proposed CATNet achieves the best performance across most tones and metrics, with an average tone precision of 87.83\% and a Cohen’s Kappa of 0.8377, surpassing all baselines. Notably, CATNet achieves best precision on two out of four tones, second-best precision on Tone 3, and maintains competitive results on Tone 2. These results demonstrate the effectiveness of our EEG-EMG architecture under audible speech conditions. 

We then choose three baseline methods with best accuracy and investigate their F-1 score and Recall in Table C1. As presented in Table C1, CATNet achieves the highest F1-score and recall across all four tones in the audio speech condition, except for Tone 1 with slight less recall score. Compared to EEGNet (83.78\% / 83.79\%) and GAT (79.54\% / 79.57\%), CATNet shows a significant performance gain, indicating its superiority in tone-level classification.
\begin{table}[H]
\centering
\resizebox{\linewidth}{!}{  
\begin{tabular}{lcccccc}
\toprule
\textbf{Method} & \textbf{Tone1} & \textbf{Tone2} & \textbf{Tone3} & \textbf{Tone4} & \textbf{Average} & \textbf{Kappa} \\
\midrule
ETE-CNN & 83.58 & 35.53 & 30.69 & 60.23 & 42.44 & 0.2325 \\
VLAAI & 94.42 & 54.64 & 64.28& 56.62 & 66.81 & 0.5578 \\
EEGNet & 98.10 & 75.88 & 87.00 & 74.11 & 83.79 & 0.7837 \\
DeepConvNet & 93.16 & \underline{83.09} & \textbf{88.79} & \underline{79.13} & \underline{86.17} & \underline{0.8154} \\
FBCSP+SVM & 87.23 & 43.61 & 41.02 & 39.27 & 52.63 & 0.3681 \\
GAT & 97.10 & 82.51 & 71.95 & 73.53 & 79.57 & 0.7270 \\
DRDA & \underline{99.21} & \textbf{91.45}  &61.24  & 58.94& 77.72 &0.7025  \\
EEG-Transformer &  97.15 & 64.93  &73.45  & 66.48 & 75.69 &0.6758  \\
CATNet & \textbf{99.34} & 82.00 & \underline{88.36} & \textbf{81.53} & \textbf{87.83} &\textbf{ 0.8377} \\
\bottomrule
\end{tabular}
}
\caption{Tone precision with baseline comparisons in audio speech condition. ETE-CNN represents End-to-End CNN. Average Precision and Kappa metrics are shown in the table. The best results in each column are highlighted in bold, and the second-best are underlined. The results are using 20 EEG channels.}
\end{table}

CATNet's also shows great stability on the most challenging tone pairs. For Tone 2, CATNet achieves an F1 score of 81.03\%. Similarly, for Tone 4, CATNet delivers the highest F1 score of 81.97\% and recall of 82.42\%, demonstrating its effectiveness in handling acoustically similar tonal variations that have historically posed difficulties for existing approaches.

In contrast, while models like GAT and DeepConvNet perform competitively on some tones, they fall short in generalization—especially on tones with subtle acoustic differences. These results collectively showcase CATNet’s effectiveness in capturing both phonetic clarity and tonal variation, setting a new benchmark for EEG-based tone classification under audio speech conditions.

\subsection{Cross Subject Results}
As shown in Table C2, CATNet achieves the highest average accuracy of 83.27\% in the audio speech condition across all ten subjects, outperforming all baseline models by a significant margin. The second-best models, GAT (77.18\%) and EEGNet (77.08\%), lag behind by over 6 percentage points, highlighting CATNet's superior generalization ability across individuals.

Moreover, CATNet demonstrates remarkable stability and consistency, achieving the highest accuracy on all 10 subjects. Notably, for subjects S5–S9, all accuracies exceed 90\%, while other baselines fails to maintain high accuracies on unseen subjects, indicating robust and scalable performance across varied participants.

\subsection{D. Model Interpretability with SHAP Values}
\renewcommand{\thefigure}{D\arabic{figure}}
\setcounter{figure}{0}
To interpret the decision-making process of our model, we employ SHAP (SHapley Additive exPlanations) values, a unified framework for explaining model predictions based on cooperative game theory~\cite{lundberg2017shap}. Our analysis focuses on two key questions: (1) How does the model shift its reliance between audible and silent speech conditions? and (2) What is the relative importance of EEG and EMG modalities?

\begin{figure}
    \centering
    \includegraphics[width=0.75\linewidth]{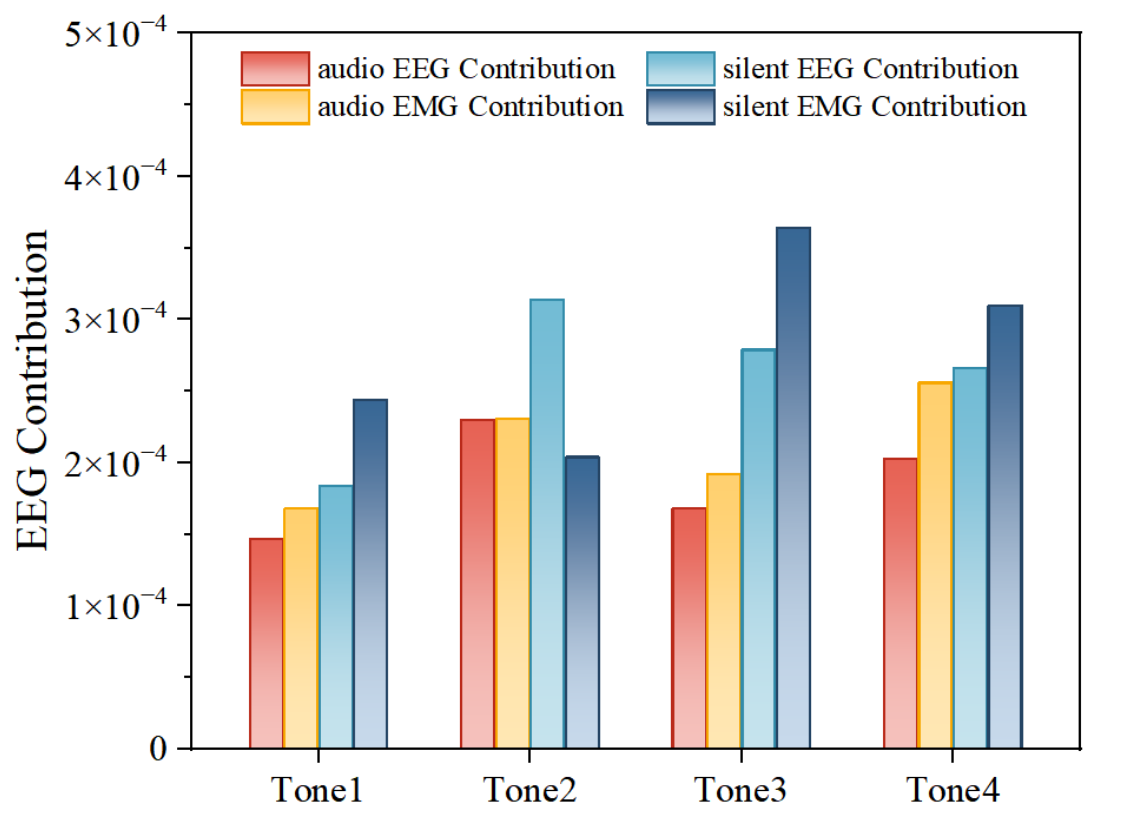}
    \caption{SHAP Value Interpreted on each tone.}
    \label{fig:enter-label}
\end{figure}

As shown in Fig. D1, EMG features generally contribute more than EEG features across both speech conditions, particularly in Tone 1, Tone 3, and Tone 4. This dominance suggests that muscle activity plays a primary role in tone classification, regardless of whether speech is audible or silent. Interestingly, Tone 2 presents an exception: EEG features contribute more than EMG features in the silent condition, highlighting the model’s increased reliance on neural signals when muscular cues are insufficient or ambiguous.

These results demonstrate that our model dynamically adjusts its modality weighting based on both speech condition and tone-specific difficulty. While EMG captures more discriminative cues in relatively easier tones (e.g., Tone 1 and Tone 3), EEG serves as a compensatory modality for more challenging tones like Tone 2, especially under silent speech scenarios.

\end{document}